\documentclass[useAMS,usegraphicx,usenatbib]{mn2e}
\usepackage{pdflscape}

\title[A census of dense cores in the Taurus L1495 cloud]{A census of dense cores in the Taurus L1495 cloud from the Herschel\thanks{{\it Herschel\/} is an ESA space observatory with science instruments provided by European-led Principal Investigator consortia and with important participation from NASA.} Gould Belt Survey}
\author[K. A. Marsh et al.]{K. A. Marsh$^{1}$\thanks{E-mail:
Ken.Marsh@astro.cf.ac.uk}, 
J. M. Kirk$^{2}$, 
Ph. Andr\'e$^{3}$,
M. J. Griffin$^{1}$, 
V. K\"onyves$^{3}$,
\newauthor P. Palmeirim$^{3,4}$, 
A. Men'shchikov$^{3}$, 
D. Ward-Thompson$^{2}$, 
M. Benedettini$^{5}$,
\newauthor D. W. Bresnahan$^{2}$,
J. Di Francesco$^{6}$, 
D. Elia$^{5}$,
F. Motte$^{3}$,
N. Peretto$^{1}$,
\newauthor S. Pezzuto$^{5}$,
A. Roy$^{3}$,
S. Sadavoy$^{7}$,
N. Schneider$^{8,9}$,
L. Spinoglio$^{5}$,
G. J. White$^{10,11}$
\\
$^{1}$School of Physics and Astronomy, Cardiff University, Cardiff CF24 3AA, UK\\
$^{2}$Jeremiah Horrocks Institute for Astrophysics and Supercomputing,
University of Central Lancashire, Preston, PR1 2HE, UK \\
$^{3}$Laboratoire AIM, CEA/DSM-CNRS-Universit\'e Paris Diderot, IRFU /
Service d'Astrophysique, C.E. Saclay, Orme des Merisiers, \\
91191 Gif-sur-Yvette, France \\
$^{4}$Laboratoire d'Astrophysique de Marseille, CNRS/INSU--Universit\'e de Provence, 13388 Marseille cedex 13, France \\
$^{5}$Istituto di Astrofisica e Planetologia Spaziali - INAF, Via Fosso del Cavaliere 100, I-00133 Roma Italy \\
$^{6}$National Research Council of Canada, Hertzberg Institute of Astrophysics,
5071 West Saanich Rd., Victoria, BC, V9E 2E7, Canada \\
Paris-Sud 11, 91405 Orsay, France \\
$^{7}$Max-Planck-Institut f\"ur Astronomie (MPIA), K\"onigstuhl 17, D-69117
Heidelberg, Germany \\
$^{8}$Universit\'e Bordeaux, LAB, UMR 5804, 33270 Floirac, France \\
$^{9}$CNRS, LAB, UMR 5804, 33270 Floirac, France \\
$^{10}$Department of Physics and Astronomy, The Open University, Milton Keynes
MK7 6AA, England \\
$^{11}$Space Science and Technology Department, CCLRC Rutherford Appleton
Laboratory, Oxfordshire OX11 0QX, England}
\begin{document}

\pagerange{\pageref{firstpage}--\pageref{lastpage}} \pubyear{2002}

\maketitle

\label{firstpage}

\begin{abstract}
We present a catalogue of dense cores in a $\sim 4^\circ\times2^\circ$
field of the Taurus star-forming region, inclusive of the L1495 cloud, 
derived from {\it Herschel\/} SPIRE and PACS 
observations in the 70 $\mu$m, 160 $\mu$m, 250 $\mu$m, 350 $\mu$m, and 
500 $\mu$m continuum bands.  Estimates of mean dust temperature and total 
mass are derived using modified blackbody fits to the 
spectral energy distributions. 
We detect
525 starless cores of which $\sim10$--20\% are gravitationally bound and
therefore presumably prestellar.
Our census of unbound objects is 
$\sim85$\% complete for $M>0.015\,M_\odot$ in low density regions 
($A_V\stackrel{<}{_\sim}5$ mag), while the bound (prestellar) 
subset is $\sim85$\% complete for $M>0.1\,M_\odot$ overall.  
The prestellar core mass function (CMF) is consistent with lognormal form, 
resembling the stellar system initial mass function, as has been reported 
previously.  All of the inferred prestellar cores lie on filamentary 
structures whose column densities
exceed the expected threshold for filamentary collapse, 
in agreement with previous reports.  Unlike the prestellar CMF,
the unbound starless CMF is not lognormal, but instead is consistent with a 
power-law form below $0.3\,M_\odot$ and shows no evidence for a 
low-mass turnover. It resembles previously reported mass 
distributions for CO clumps at low masses ($M\stackrel{<}{_\sim}0.3\,M_\odot$).
The volume density PDF, however, is accurately lognormal except at high 
densities. It is consistent with the effects of self-gravity on magnetized
supersonic turbulence. The only significant deviation from lognormality
is a high-density tail which can be attributed unambiguously to 
prestellar cores.
\end{abstract}


\section{Introduction}
While much progress has been made in determining the form of the stellar initial
mass function (IMF) \citep{chab03,kroupa13}, a detailed explanation of its 
form and possible environmental dependence must hinge on an understanding of 
the nature and evolution of the precursor cold dense cores. 
These cores are best observed in the submillimetre regime corresponding to 
their peak thermal dust emission; observations made with
ground-based telescopes have previously helped to establish important 
links between the IMF and the prestellar core mass function (CMF)
(e.g. \citealt{motte98}). With the 
advent of {\it Herschel\/}, however,
these cores can now be studied with much higher sensitivity and hence
the CMF can be probed to lower masses than before. 

One of the major goals
of the {\it Herschel\/} Gould Belt Survey (HGBS) \citep{and10} is to 
characterise the prestellar CMF over the Gould Belt. 
This survey covers
15 nearby molecular clouds which span a wide range of star formation
environments.  
A principal product of the HGBS will be a 
catalogue, for each cloud, of all detected dense cores and protostellar objects.
The first installment covers the Aquila molecular cloud;
preliminary results were reported by 
\citet{kon10} and the full Aquila catalogue is now available
\citep{kon15}. The present paper represents the next installment, which
consists of a list of detected dense cores in a portion of the Taurus
Molecular Cloud containing L1495. This catalogue
is available for download in the electronic edition of
this paper. A portion is shown in Appendix B to illustrate the format and
content. 

The Taurus cloud represents a nearby region of predominantly non-clustered 
low-mass star formation, at an estimated distance of 140 pc \citep{elias78},
in which the stellar density is relatively low and objects can be studied
in relative isolation.  Its detailed morphology at {\it Herschel\/} wavelengths
is discussed by \citet{kirk2013}. Overall, it is dominated by two long
($\sim 3^\circ$), roughly parallel filamentary structures, the northern
of which is the larger.  Early results from {\it Herschel\/}
regarding the filamentary properties have been reported by \cite{palm13}.  
The catalogue presented here covers the L1495 cloud located in the
western portion of the northern filamentary structure,
designated as N3 in \cite{kirk2013}.

\section[]{Observations}

The observational data on which the present catalogue is based consists of a 
set of images of 
the L1495 cloud in the Taurus star-forming region, made as part
of the HGBS\footnote{http://gouldbelt-herschel.cea.fr} \citep{and10}.  The data were taken using
PACS \citep{pog10} at 70 $\mu$m and 160 $\mu$m and SPIRE \citep{griffin10} 
at 250 $\mu$m, 350 $\mu$m, and 500 $\mu$m in fast-scanning 
(60 $''{\rm s}^{-1}$) parallel mode.  The North-South scan 
direction for L1495 was split into two observations taken on 12 February 
2010 and 7 August 2010, while the East-West cross-scan direction was observed
in a single run on 8 August 2010.  The corresponding {\it Herschel\/} 
Observation IDs are 1342202254, 1342190616, and 1342202090.
An additional PACS observation (Observation ID 1342242047) was taken
on 20 March 2012 in the orthogonal scan direction to fill a gap found in the
previous PACS data. 

Calibrated scan-map images at the PACS wavelengths were produced in 
the HIPE Version 10 pipeline \citep{ott2010} using Scanamorphos 
Version 20 \citep{rous13}. 
The SPIRE images were produced in the same pipeline using the
``naive" map-making procedure and the destriper module.
The arbitrary sky offset values were corrected
by comparison with Planck and 
{\it IRAS\/} data using a procedure similar to that employed by \citet{bern10}.
The estimated offset values\footnote{The adopted sky median values 
for the Taurus N2/N3 regions are: 4.27, 69.2, 47.0, 29.0, and 13.4 
MJy sr$^{-1}$ at wavelengths of 70 $\mu$m, 160 $\mu$m, 250 $\mu$m,
350 $\mu$m, and  500 $\mu$m, respectively.} 
for SPIRE correspond well to the ones
determined from the zero-point correction task in HIPE.
The spatial resolution values (FWHM beamsizes) of these maps are approximately
$8.5''$, $13.5''$, $18.2''$, $24.9''$, and $36.3''$ in the 70 $\mu$m, 
160 $\mu$m, 250 $\mu$m, 350 $\mu$m, and 500 $\mu$m wavelength bands, 
respectively. In the PACS bands, however, the effective beams resulting
from fast scanning are somewhat elongated, at
$\sim6''\times12''$ (70 $\mu$m) and $\sim12''\times16''$ (160 $\mu$m).
The present catalogue is based on a rectangular subregion
of size $\sim4^\circ\times2^\circ$, whose long axis is aligned approximately 
with the axis of the long filamentary structure at a position
angle of $-52^\circ$.

\section[]{Source detection and classification}

The production of the catalogue of cores for all of the various 
star-forming regions 
within the Gould Belt represents a parallel effort by a number of groups.
To ensure uniformity, all groups are adhering strictly to the 
procedure specified in detail by \citet{kon15}. 
The identification of starless cores is then essentially a two-step
process:
\smallskip

\noindent{\it Step 1: Detection:}
Source detection is carried out using the {\it getsources\/} algorithm
\citep{men12} as provided by the Nov.\ 2013 major release (Version 
v1.140127). Starless core candidates are detected while making simultaneous
use of images in the four {\it Herschel\/} wavebands in the 
160--500 $\mu$m range, the 160 $\mu$m image being a temperature-corrected
version (see \citealt{kon15} for details). In addition, a column density
map is used as if it were an additional waveband, the purpose being to
give extra weight to regions of high column density. This
map\footnote{Subsequently referred to as the ``high-resolution"
column density map.}, with an effective spatial resolution of $18.2''$
corresponding to the 250 $\mu$m beam, is obtained using the technique 
described in Appendix A of
\citet{palm13}. 
The output of the detection step consists only of sources
deemed ``reliable" on the basis of the 
{\it getsources\/} ``global goodness" parameter, which combines the global 
detection significance, the combined signal-to-noise ratio ($S/N$) 
over all bands, and source geometry knowledge. 

As discussed by \citet{kon15}, the detection of cores is carried out 
separately from that of protostellar objects, the latter
being based on a second run of {\it getsources\/} in which detection
is based on the 70 $\mu$m data only.  The {\it Herschel\/} results for
protostellar objects in this region will be discussed separately
(J. M. Kirk et al., in preparation).
\clearpage

\noindent{\it Step 2: Classification:}

Classification as a starless core candidate requires that all of the following conditions 
be met:
\begin{enumerate}
  \item column density detection significance greater than 5 (where 
significance in this case refers to detection at the relevant single spatial 
scale, as defined by \citealt{men12});
  \item column density $(S/N)_{\rm peak} > 1$ at $18.2''$ resolution;
  \item global detection significance over all wavelengths \citep{men12} 
greater than 10;
  \item flux detection with significance greater than 5 for at least two
wavelengths between 160 $\mu$m and 500 $\mu$m;
  \item flux measurement with $S/N>1$ in at least one band between
160 $\mu$m and 500 $\mu$m for which the monochromatic detection significance
is greater than 5;
  \item detection significance less than 5 for the 70 $\mu$m band {\it or \/}
  70 $\mu$m peak flux per pixel less than zero {\it or\/}
  source size (full width at half maximum, FWHM) at 70 $\mu$m 
  larger than 1.5 times the 70 $\mu$m beam size;
  \item source not spatially coincident with a known galaxy, based on
comparison with the NASA Extragalactic Database (NED) or with a known young
stellar object (YSO) based on the published list of Widefield Infrared Survey
Explorer (WISE) detections \citep{reb11};
  \item source appears real based on visual examination of the SPIRE images and
column density maps.
\end{enumerate}

\begin{figure}
\includegraphics[width=90mm]{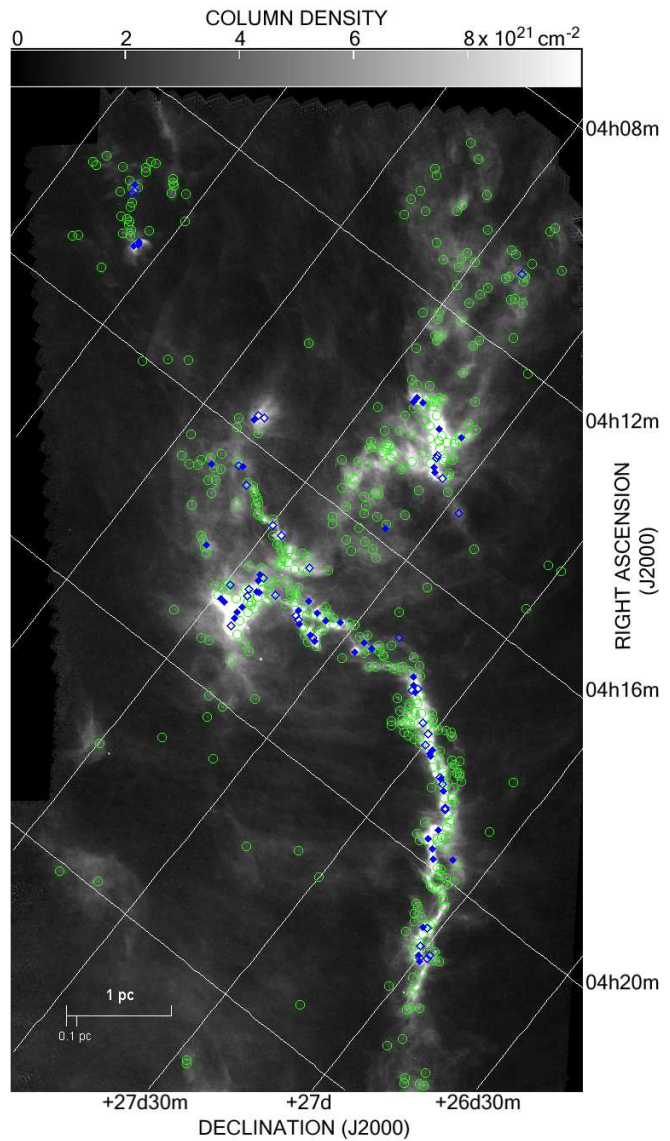}
 \caption{Column density map of the region studied. The spatial resolution
is approximately $18.2''$ and the column density units are hydrogen molecules
per cm$^2$. The overplotted symbols represent the locations of detected cores 
(green circles = unbound starless, filled blue diamonds = prestellar, open
blue diamonds = candidate prestellar). The peak column
 density is $5.5\times10^{22}\,{\rm cm}^{-2}$, although the display scale is
truncated at $1.0\times10^{22}\,{\rm cm}^{-2}$ to improve the visibility of
faint features.}
 \label{fig1}
\end{figure}

Imposition of the last criterion is facilitated by examination of the
``summary card" for each source, included as one of the on-line 
catalogue products\footnote{These ancillary data consist of 
the full catalogue, a table of derived parameters, and the set of summary 
cards. They are available in the on-line version of this paper, and from the
HGBS website at http://gouldbelt-herschel.cea.fr/archives. Also available
from the latter link are all of the HGBS data products, which include
{\it Herschel\/} images and the column density and dust temperature maps.}.
It includes a cutout column density image of the source, 
upon which is superposed the 
elliptical 50\% contour as estimated by {\it getsources\/}.  To provide some
additional confidence in the reality of the extracted source, the subimage is 
also superposed with a source ellipse estimated by an entirely different
source extraction technique, namely the Cardiff Source-finding
AlgoRithm ({\tt CSAR}) discussed in Appendix B of \citet{kirk2013}. The
latter was run on the high-resolution ($18.2''$) column density map.
The above procedure resulted in 525 objects classified as starless 
cores, of which 46\% were detected by {\tt CSAR}\footnote{A source
detected by {\it getsources\/} is considered to have a CSAR counterpart if the
CSAR position falls within the 50\% elliptical contour of the high-resolution
column density map as measured by {\it getsources\/}.}.
Their locations are shown in Fig. \ref{fig1}
relative to a column density image of the field produced by the
\citet{palm13} technique.

Although most of the sources in Fig. \ref{fig1} are clearly associated with 
filamentary structure in L1495, we cannot rule out the possibility
that a few of them may be extragalactic contaminants. Some extragalactic
sources may, in fact, have been missed in the NED as a result of 
foreground confusion by Taurus itself. To assess this possibility we
note that the estimated rms level of extragalactic contaminants is 
$<10$ mJy in {\it Herschel\/} bands between 160 $\mu$m and 500 $\mu$m
\citep{gar14}, more than an order of magnitude lower than that
of 99\% our detected cores. Nevertheless, some extragalactic sources may 
be present in the remainder.

\section[]{Estimates of core masses and temperatures}

Estimates of the core masses and dust temperatures were obtained for each 
source by fitting a modified blackbody spectrum to the observed spectral
energy distribution (SED) constructed from the set of 
{\it getsources\/} flux densities in the wavelength 
range 160--500 $\mu$m.  The SED point at each wavelength represents the 
flux density integrated over the entire source, and overlapping sources
are deblended as described by \citet{men12}. The flux densities used
in SED fitting are presented in Table B1.
For this exercise, sources were assumed to be 
isothermal and have a wavelength variation of opacity of the form
\citep{hild83,roy2014}:
\begin{equation}
    \kappa(\lambda) = 0.1\,(300/\lambda)^2\qquad{\rm cm}^2\,{\rm g}^{-1}
    \label{eq1}
\end{equation}
where $\lambda$ is the wavelength in $\mu$m. 
The fitting procedure used was the same as that described by \cite{kon15},
and involved a weighted least squares fit in which each SED flux was
weighted by the inverse square of its measurement uncertainty, as estimated
by {\it getsources\/}. Those uncertainties are dominated by errors 
associated with background subtraction. 

The fits were based on an
assumed distance of 140 pc.
Following \citet{kon15}, the 
500 $\mu$m flux was
excluded from the fit in the small minority of cases in which this flux
exceeded that at 350 $\mu$m, as a precaution against confusion contamination
at the longer wavelength. 

Some of the detected cores can be expected to be gravitationally bound,
and thus likely to go on to form stars. We therefore refer to the members of
this subset as prestellar cores, and identify them by modelling the 
sources as Bonnor-Ebert (B-E) spheres \citep{bonnor1956,ebert1955}. 
For a source of outer radius $R$, the B-E
critical mass \citep{bonnor1956} is given by:
\begin{equation}
M_{\rm BE}=2.4\,R\,a_s^2/G
\label{eq4}
\end{equation}
where $G$ is the gravitational constant and 
$a_s$ is the isothermal sound speed. The latter is given by
$a_s = (kT/\mu)$ where $k$ is Boltzmann's constant and $T$ is the gas
temperature which we equate with the estimated dust temperature.

The correspondence between $R$ and the source size estimated using
{\it getsources\/} has been investigated by \citet{kon15}. 
Briefly, simulations of {\it getsources\/} extractions on a set of 
B-E source models show that the {\it getsources\/} FWHM values
provide good measurements of the outer radii of critical B-E spheres
(with an accuracy better than 10-20\%) if spatially well resolved.
For marginally-resolved cores, the {\it getsources\/}
FWHM estimate provides a 
reliable estimate of the true B-E radius after convolution with the beam.
For cores with extended power-law wings, however, the source
sizes are not measured as reliably by the current
version of {\it getsources\/}. 
Based on these considerations we take $R$ as the geometric mean of the 
source FWHM major and minor axes as estimated by
{\it getsources\/} from the 
column density map made at $18.2''$ resolution, after simple beam
deconvolution (quadrature subtraction of the beam FWHM).  

Following \citet{and07} and \citet{kon15}, we then classify a detected starless 
core as ``prestellar" if its estimated mass exceeds $M_{\rm BE}/2$ 
(cf. \citealt{bert92}). The
factor of 2 corresponds to the value of
the virial ratio below which an object would be regarded as gravitationally 
bound, taking $M_{\rm BE}$ as a proxy for the virial mass.
Since the mass threshold depends linearly on $R$, misclassification is likely
for many marginally-resolved sources. In order to avoid rejecting genuine
prestellar cores in that category entirely, we provide an additional list
of ``candidate prestellar cores"  which, although lacking sufficient mass
for robust classification as prestellar, nevertheless remain 
plausible candidates. We select them
by lowering the threshold value of $M/M_{\rm BE}$ from 0.5 (well-resolved
cores) down to 0.2 (marginally-resolved cores). The actual functional 
dependence of the threshold, $t$, on angular source size has been estimated, 
in the case of the Aquila region, by \citet{kon15} using simulations. They
obtained 
$t = 0.2\,(\theta_{\rm source}/\theta_{\rm beam})^{0.4}$ 
subject to $0.2 \le t \le 0.5$,
where $\theta_{\rm source}$ and $\theta_{\rm beam}$ represent the 
FWHMs of source and beam in the high-resolution
column density map.
Our Taurus simulations, conducted in connection with the completeness estimates 
described in Appendix A, suggest that the latter expression is appropriate
here also. 

\begin{figure}
\hspace*{-0.2cm}\includegraphics[width=88mm]{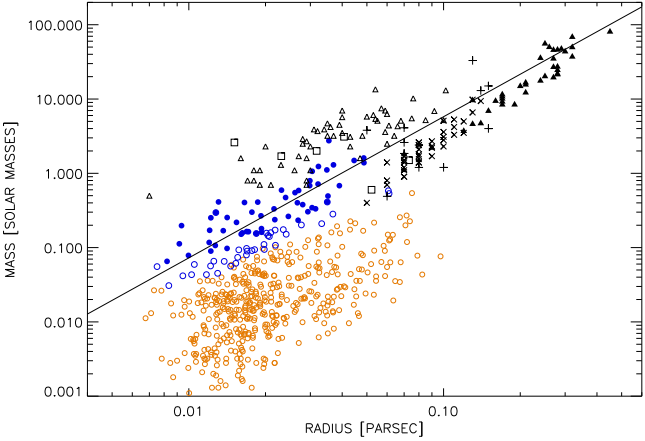}
 \caption{The mass-radius correlation for cores in Taurus. The coloured
symbols represent our estimates of core mass as a function of deconvolved radius
for the L1495 sources classified as starless cores.  Blue filled circles,
blue open circles, and gold open circles represent prestellar cores,
candidate prestellar cores, and gravitationally unbound starless cores,
respectively.
Black symbols represent previously published Taurus data, as follows:
{\it Squares\/}: Observations at 850 $\mu$m \citep{kirk2005}; ``$\times$": 
Extinction-based measurements of cores in L1495 \citep{schmalzl2010}; 
{\it Filled triangles\/}: Estimates based on C$^{18}$O observations 
\citep{onishi1996}; 
{\it Open triangles\/}: Estimates based on H$^{13}$CO$^+$ observations 
\citep{onishi2002}; 
``$+$": Estimates based on NH$_3$ observations 
\citep{bens89}. Also shown for comparison (solid line) is the mass-radius 
relation for clumps in molecular clouds \citep{lars81}, 
represented by $M/M_\odot =460\, R_{\rm [pc]}^{1.9}$.} 
 \label{fig2}
\end{figure}

Fig. \ref{fig2} shows a plot of estimated mass, $M$, versus radius, $R$.
The plotted points
are distinguished based on their gravitational status, i.e., bound,
possibly bound (candidate prestellar), or unbound.
Also plotted
for comparison are previous Taurus data from the literature.
Clearly, the masses and radii are well correlated and are
consistent with the power-law
relation found from previous studies of dense cores and clumps in molecular
clouds (e.g., \citealt{lars81,kauff10}).  They are also consistent with
extinction-based results for L1495 \citep{schmalzl2010},
C$^{18}$O \citep{onishi1996}, H$^{13}$CO$^+$ \citep{onishi2002}, and
with the \citet{kirk2013} results for other parts of the Taurus molecular cloud
(the Barnard 18 and L1536 clouds) based on {\it Herschel\/} data, and
with NH$_3$ data \citep{bens89}.

\begin{figure}
\hspace*{-0.2cm}\includegraphics[width=88mm]{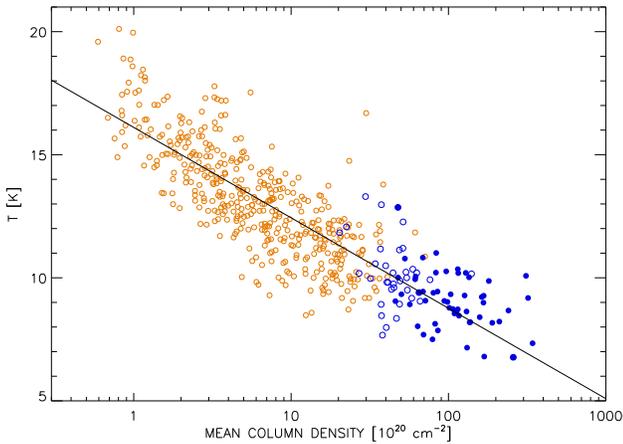}
 \caption{Estimated mean dust temperature plotted as a function of 
mean core column density, $N_{{\rm H}_2}=M/(\pi R^2\mu m_{\rm H})$, where $\mu$
is the mean molecular weight (taken as 2.8) and $m_{\rm H}$ 
is the mass of a hydrogen atom. Blue filled circles, blue open circles,
 and gold open circles represent prestellar, candidate prestellar, and
unbound starless cores, respectively.
respectively. The straight line 
represents a linear regression fit to all of the points, and corresponds 
to a $T$[K] versus $N_{{\rm H}_2} [10^{20}\,{\rm cm}^{-2}]$ relation of the form
$T = -3.6 \log_{10} N_{{\rm H}_2} + 16.1$. This correlation can be expected to
be region-dependent due to its likely dependence on the local ISRF.}
 \label{fig3}
\end{figure}

Fig. \ref{fig3} shows a plot of dust 
temperature versus mean column density and indicates that these quantities
are negatively correlated. Such a correlation might be expected based on a 
model in which the
dust is heated by the interstellar radiation field (ISRF). Large
column densities then provide more shielding, resulting in cooler dust
\citep{evans01,stam07,roy2014}. 
But could this correlation be an artifact, induced 
by a degeneracy between mass and temperature in the SED measurement model?  
Such a degeneracy is certainly possible in cases of limited wavelength coverage,
particularly if the observed wavelengths are all on the
Rayleigh-Jeans side of the Planck peak for the particular dust temperature.
For the present observations, however,  the 
{\it Herschel\/} wavelengths bracket the peak of the modified blackbody function for 
all of the core temperatures likely to be encountered. It is true that
temperature gradients within cores can also give rise to mass biases (and hence
column density biases) when using simple SED fitting,
but these biases are typically at the $\stackrel{<}{_\sim}30$\% level (see, 
for example, \citealt{roy2014,kon15}) and not sufficient to  
produce the correlation apparent in Fig. \ref{fig3}.
Furthermore, the latter correlation remains even after full account
is taken of the effects of radial temperature gradients \citep{marsh2014}.
We thus conclude that the correlation is real, and consistent with 
the ISRF heating model \citep{stam07}.

\begin{figure}
\hspace*{-0.2cm}\includegraphics[width=88mm]{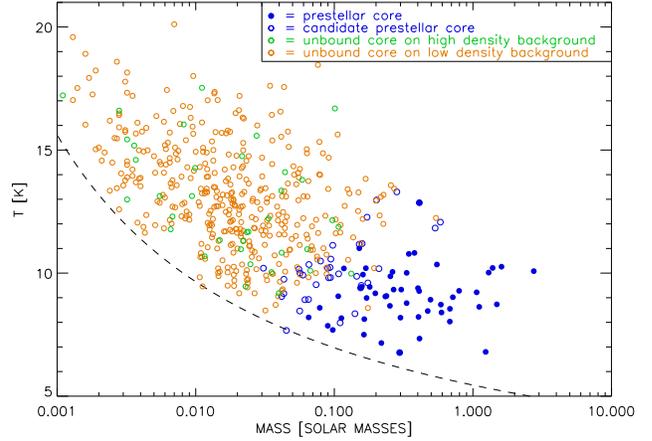}
 \caption{Estimated mean dust temperature plotted as a function of 
estimated core mass. Bound (prestellar) cores and prestellar core candidates
are plotted as filled and open blue circles, respectively.
Unbound cores are represented by open circles and are
divided into those seen against high column density backgrounds
characterised by $A_V>7$ mag (green symbols), and more diffuse backgrounds 
(gold symbols). The dashed line represents the detection limit, i.e., the
lowest detectable mass for a given core temperature.}
 \label{fig4}
\end{figure}

The temperature versus mass plot of Fig. \ref{fig4} shows that these
quantities are at least partially correlated
in the sense that the more massive cores ($M\stackrel{>}{_\sim}1\,M_\odot$)
have the lowest temperatures. There is, however, no information as to whether 
the converse is true; the detection sensitivity cutoff prevents us from
knowing whether all low mass cores have high temperatures.
Therefore, although this figure
indicates that, within the detected population, unbound starless cores
have systematically higher temperatures than prestellar cores, we cannot
rule out the existence of a population of cool starless cores. This issue
is discussed further in Appendix A.

\section[]{Overall starless core mass function}

\begin{figure}
\hspace*{-1.0cm}\includegraphics[width=100mm]{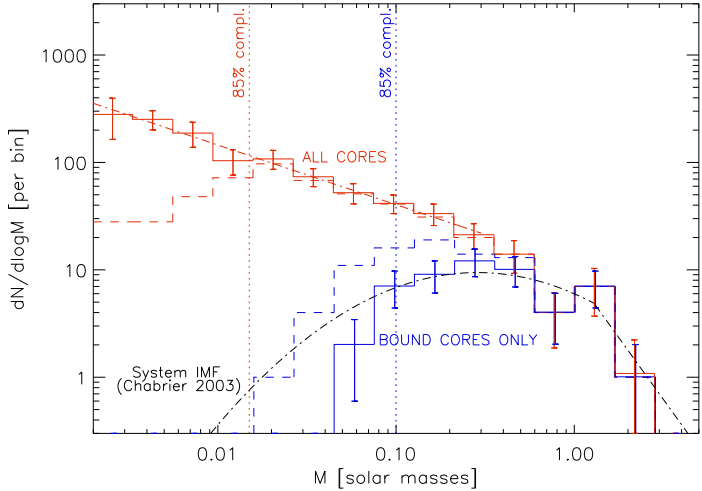}
 \caption{Estimated core mass function.  The dashed red histogram 
shows the CMF for the complete set of detected cores, uncorrected for the 
incompleteness of the observations.  Applying an 
approximate correction for incompleteness, based on the results in Appendix A, 
yields the solid red histogram.  The dash-dot red line represents a function
of the form $dN/d\log M \propto M^{-0.55}$, obtained by linear regression of
the logarithmic variables in the mass range $M/M_\odot<0.3$.
The solid blue histogram shows the 
gravitationally bound (i.e., prestellar) subset, while the dashed blue
histogram represents the subset consisting of some additional 
{\it candidate\/} prestellar objects.
In neither case has any correction been applied
for incompleteness. The vertical dotted lines represent the 85\% completeness
limits for unbound starless cores (red) and prestellar cores (blue), as
discussed in Appendix A.
Also shown for comparison (black dash-dot curve) is a scaled version of
the \citet{chab03} system IMF in which masses have been scaled upwards by a
factor of 1.3.}
 \label{fig5}
\end{figure}

The dashed red histogram in Fig. \ref{fig5} represents the uncorrected mass 
distribution 
for the sample as a whole, without regard for inferred gravitational 
status (bound or unbound).
One feature of this distribution is an apparent decrease in source counts
for masses below $\sim0.01\,M_\odot$, in the
mass regime dominated by unbound starless cores. However,
comparison with the completeness curve for unbound cores (Appendix A)
suggests that this behaviour may be an effect of incompleteness.
Although the detailed form of the completeness curve is 
subject to uncertainty as discussed in Appendix A, it is nevertheless
instructive to apply it as a correction to the raw CMF to 
get an indication of how the true starless CMF may 
look. The result is shown by the solid red histogram in Fig. \ref{fig5}.
We conclude that there is no evidence for a low-mass turnover 
in the starless CMF in the range surveyed by {\it Herschel\/} and
that the overall starless CMF probably continues to rise towards even
lower masses.  This result means that
the apparent turnovers in previous estimates of the Taurus
CMF (for example, the submillimetre-based estimate of \citealt{sad10}
and the extinction-based estimate of \citealt{schmalzl2010})
were due to the incompleteness of those surveys. In the case of the
\citet{sad10} survey, some of the incompleteness in the overall starless
core population can be attributed to the fact that ground-based
850 $\mu$m observations preferentially select the {\it densest\/} objects, 
i.e., those most likely to be prestellar, for reasons discussed by
Ward-Thompson et al. (in preparation).  In any event, the {\it Herschel\/}
observations have served to extend the overall starless CMF, as derived 
from dust emission, by an order of magnitude in mass below that of previous 
estimates.

In the mass range dominated by unbound cores 
($M\stackrel{<} {_\sim}0.3\,M_\odot$), the CMF in Fig. \ref{fig5} is 
consistent with power-law 
behaviour, i.e., $dN/d\log M\propto M^\alpha$. 
A linear regression with respect to the logarithmic quantities then yields
the maximum likelihood estimate $\alpha = -0.55\pm0.07$.
This is consistent with the observed power-law distribution of CO clumps,
for which a representative value of $\alpha$ is
$-0.7\pm0.1$ based on the collective data for a number of
star-forming regions \citep{kram98}.
Such clumps are the relatively low density structures previously detected
in CO emission, the majority of which are gravitationally unbound
\citep{blitz93,kram98}. 
As with the {\it Herschel\/} unbound cores, the CO clump distribution
shows no low-mass turnover \citep{kram98}. Clumps identified by various 
isotopologues of CO all show fairly similar slopes in their mass distributions
at low masses. 
For example, observations of the
Perseus molecular cloud give $\alpha$ values of $-0.36\pm0.08$ 
($0.1\!<\!M/M_\odot\!<\!3$) and 
$-0.4\pm0.2$ ($0.03\!<\!M/M_\odot\!<\!5$) 
for C$^{18}$O(3-2) and $^{13}$CO(3-2) \citep{curt09}, respectively, both
of which are consistent with our estimate based on dust emission for
the mass ranges in common.
The maps of C$^{18}$O are, in fact, strikingly similar to those in the
850 $\mu$m continuum \citep{curt10}, suggesting that similar densities of 
material are being traced.  It seems likely, therefore, that
our unbound starless cores represent the dust-emission counterparts
of the C$^{18}$O clumps. 

The \citet{curt09} observations also revealed a population of
C$^{18}$O clumps with no 850 $\mu$m counterparts. These were low-mass
objects which subsequent virial analysis showed to be gravitationally
unbound. Similar behaviour was found in $\rho$ Oph by 
\citet{white2015}. Their lack of detection at 850 $\mu$m was due, at least 
in part, to the fact that the 850 $\mu$m observations were not sensitive
to masses below about 0.1 $M_\odot$.  The {\it Herschel\/} data will
allow such analyses to be extended to lower masses and provide
further elucidation of the nature of gravitationally unbound cores.

\section{The prestellar core subset}

We find that 52 of our 525 starless cores fulfil the requirements for 
robust classification as gravitationally bound, i.e., prestellar, based
on the criteria listed in Section 4. An additional 38 cores may be
regarded as candidate (non-robust) detections of prestellar cores based
on the somewhat more relaxed criteria specified in that section. The total 
number of prestellar cores in the starless sample is therefore somewhere
in the range 52--90, representing $\sim10$--20\% of the starless core sample.
This represents a significant
difference with respect to the Aquila region, for which the bound
fraction of {\it Herschel\/} cores is $\sim60$\% \citep{kon15}.
The CMFs of the robust and candidate subsets of bound cores are shown by the 
solid and dashed blue histograms, respectively, in Fig. \ref{fig5}. 
There may well be even more prestellar cores whose masses fall below
the completeness limit of $\sim\!\!0.1\,M_\odot$.
Such a possibility is raised by the previous interferometric detection 
of a pre-brown dwarf candidate in Ophiuchus/L1688 \citep{and12} 
whose mass, $\sim0.02\,M_\odot$, falls well below the {\it Herschel\/} 
completeness limit.

Above the completeness limit, the prestellar CMF is consistent with
lognormal form and resembles the
\citet{chab03} estimate of the stellar system IMF, consistent with
previous findings for other star-forming regions \citep{alves07,simp08,and10}. 
Those studies did, however, find that to reconcile the CMF with the IMF,
scaling was required on the mass axis, and the scaling factor was interpreted
in terms of the star formation efficiency based on a simple model in which
each core gives rise to one stellar system. 
We estimate the scaling factor by multiplying the mass axis of the 
\citet{chab03} IMF by various factors, and selecting the value which minimises
the weighted sum of squares of residuals with respect to the observed CMF.
The weighting used is inverse-variance, based on the 
Poisson error bars of the observed CMF values. The optimal scaling
factor is then 1.3, but
given the large statistical uncertainties due to the 
relatively small number of prestellar cores, the data are consistent with
values ranging from $\sim\!1$ to $\sim\!2$.
Hence there is no evidence for
star formation efficiencies which differ significantly from the value
$\sim40$\% reported for Aquila \citep{kon15}. These conclusions are unchanged
if the \citet{chab03} system IMF is replaced by those of \citet{kroupa13}.

Errors in our estimated CMF can arise from a number of effects. As mentioned
earlier, the non-isothermal structure of cores has an effect on masses
estimated by the simple SED fitting procedure used here.
As discussed by \citet{kon15}, the masses can be underestimated by
factors ranging from $\sim20$\%  to
$\sim2$ depending on the column density (and hence, indirectly, the mass)
of the core.
Consequently, the number of cores classified as prestellar is reduced. 
The masses may be further underestimated due to
the underestimation of outer radii for cores with developed power-law wings. 
More detailed modelling of the temperature and density structure would
alleviate these effects, but in the meantime we cannot say conclusively that
the prestellar CMF in L1495 is of lognormal form, but only that it is consistent
with that form given the current level of uncertainty. 

\subsection[]{Relationship to filaments}
{\it Herschel} observations have provided strong indications of an 
evolutionary link between filaments and cores \citep{and10,and14}, and so
it is of interest to ask what fraction of prestellar cores lies on filaments.
In this regard \citet{poly13} found that 71\%
of prestellar cores in the L1641 cloud of Orion A were located on filaments
with $A_V>5$ mag,
leaving a substantial fraction that were not. This may be compared
with the early estimate of $>60$\% for the fraction on supercritical
filaments in Aquila \citep{and10} and the more recent (and more robust)
estimate of $75^{+15}_{-5}$\% by \citet{kon15}.

There is no common consensus in the literature regarding the definition of
a filament. Most definitions are observationally based and therefore quite
subjective.  To provide a context for subsequent discussion,
we adopt as a working definition the one given by
\citet{and14}, i.e., ``any elongated ISM structure with an aspect ratio larger
than $\sim5$--10 that is significantly overdense with respect to its
surroundings," but qualify it further by specifying that the elongated 
structure must be visible directly in a column density image 
that has not been spatially filtered.

\begin{figure}
\includegraphics[width=84mm]{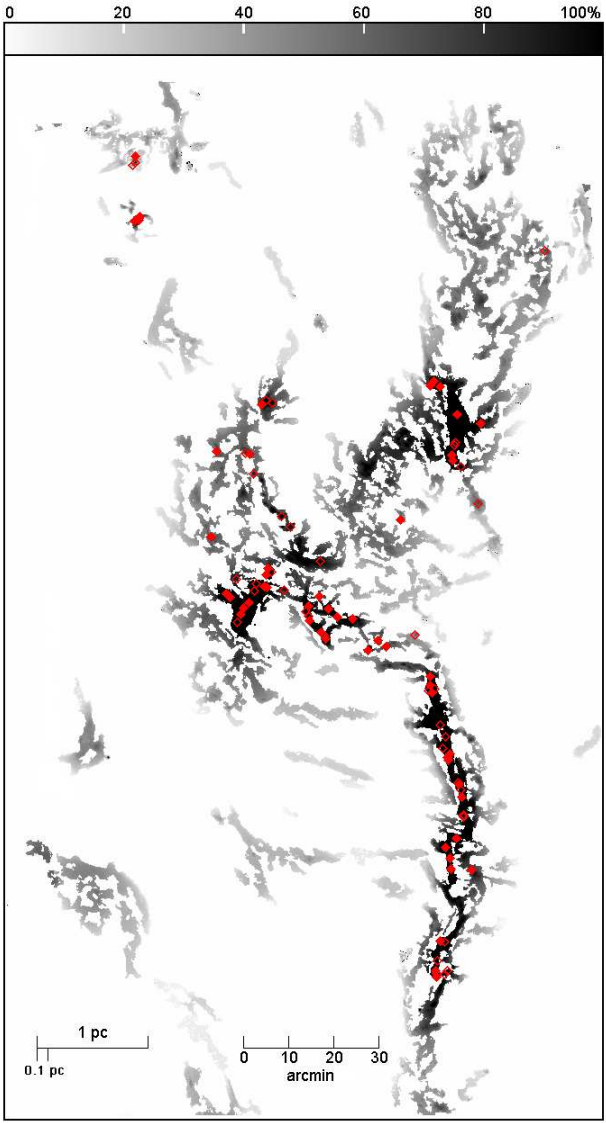}
 \caption{The locations of prestellar cores (filled red diamonds) and
additional prestellar core candidates (open red diamonds) with respect to
filamentary structure, shown with the same field of view as for 
Fig.~\ref{fig1}. The image has been rotated such that equatorial north
is $52^\circ$ anticlockwise from vertical.
The filamentary structure represents a limited-scale reconstruction, up to a 
transverse spatial scale of 0.1 pc, obtained using 
the {\it getfilaments\/} algorithm \citep{men13}.
The greyscale values correspond to the
{\it total\/} (unfiltered) background line densities within the boundaries 
of the reconstructed features, based on an assumed characteristic
filamentary width of 0.1 pc, and are truncated at an upper value 
corresponding to 16 $M_\odot\,{\rm pc}^{-1}$ (100\% on the greyscale). 
As a result, the portions in
black have a mass per unit length in excess of the approximate threshold
for cylindrical stability.}
 \label{fig6}
\end{figure}

To investigate the situation for L1495, we compare the locations
of our inferred prestellar cores with the background filamentary structure 
extracted using
the recently developed {\it getfilaments\/} algorithm \citep{men13}.
The latter is similar to {\it getsources\/} in that it involves
filtering at a range of spatial scales. 
Fig. \ref{fig6} shows the results obtained from the L1495
column density map. It represents a reconstruction
of elongated structures up to a transverse spatial scale of
$145''$ which, at 140 pc, corresponds to the 0.1 pc characteristic 
filamentary width found in previous studies \citep{arzo11,palm13}. 
Note that the spatial filtering has suppressed the broad power-law
wings revealed by those studies. The wings actually extend considerably 
further out than the 0.1 pc corresponding to the FWHM of the cross-sectional 
profile.
The spatial filtering does, however, serve to show the smaller scale
features superposed on the overall filament structure.
Some of these features are in the form of 
threadlike structures which correspond very well to the velocity coherent
``fibres" identified by \citet{hac13} in C$^{18}$O (see Fig. 2 of
\citealt{and14}). We will use the term ``filamentary structure" to
include both the large-scale filaments and the smaller-scale fibres.

\begin{figure}
\includegraphics[width=88mm]{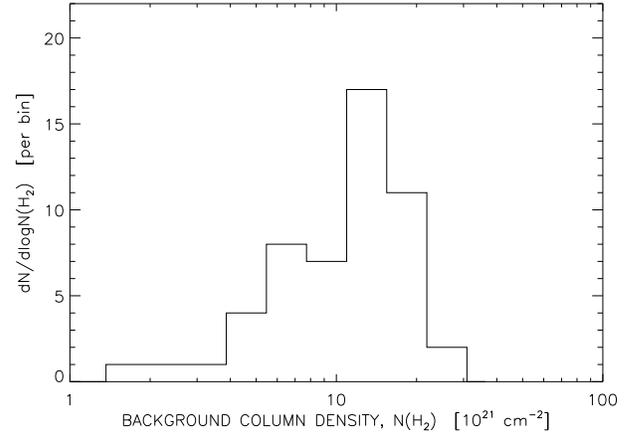}
 \caption{The distribution of filamentary background column densities 
associated with prestellar cores.}
 \label{fig7}
\end{figure}

Superposed on Fig. \ref{fig6} are the core locations. We find that 100\% of 
the prestellar cores are located on filamentary structure with $A_V \ge 5$ mag,
as opposed to 40\% for the starless core sample as a whole.
A similarly strong correlation was found between
YSO locations and filamentary structure in Taurus by \citet{doi15}.
Fig. \ref{fig7} shows a histogram of
total background column density values at the 
locations of the prestellar cores. It is prominently peaked at
$\sim 1.3\times10^{22}\,{\rm cm}^{-2}$ and indicates that 90\% of the 
column densities exceed $5.9\times10^{21}\,{\rm cm}^{-2}$. 
The latter, which may be interpreted as an approximate lower threshold
for star formation, is consistent with the value
$\sim7\times 10^{21}\,{\rm cm}^{-2}$ for Aquila \citep{and14,kon15} and also
with the threshold found by \citet{gold08} for Taurus young stars.
If we take 0.1 pc as the mean filamentary width, then the 
estimated line masses of filamentary structure in the immediate vicinities of
prestellar cores throughout L1495 are in the range 
$32\pm15$ $M_\odot\,{\rm pc}^{-1}$. Our mean value is about a factor of two
larger than the value 15 $M_\odot\,{\rm pc}^{-1}$ found by \citet{hac13} 
using N$_2$H$^+$ observations, and the value 
17 $M_\odot\,{\rm pc}^{-1}$ found by \citet{schmalzl2010} based on 
near-infrared extinction measurements. The reason for the discrepancy
is not clear, but it is significant that all three estimates are
greater than or equal to the theoretical value of
16 $M_\odot\,{\rm pc}^{-1}$ for an isothermal cylinder in pressure equilibrium
at 10 K \citep{inut97}. This suggests that all of the prestellar
cores in L1495 are located in supercritical filaments, in agreement
with the HGBS findings in Aquila \citep{and10,kon15}. 
We note that none of the above estimates 
includes the contribution of the extended power-law wings whose inclusion 
further increases the estimated line mass. For example, the 
line mass integrated over the full filamentary cross section of the
B211 filament is estimated to be 54 $M_\odot\,{\rm pc}^{-1}$ \citep{palm13}.

\subsection[]{Relationship to unbound starless cores}

\begin{figure}
\includegraphics[width=88mm]{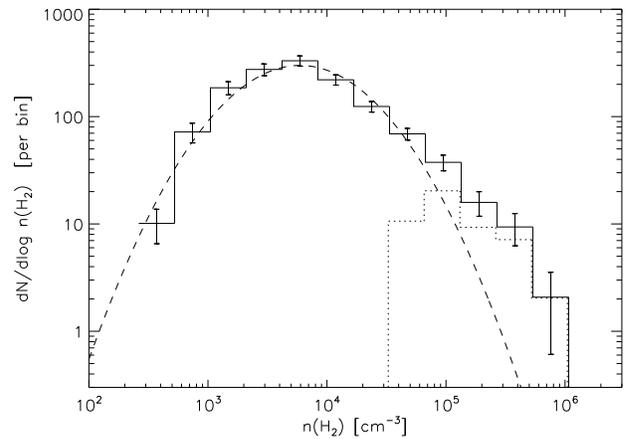}
 \caption{Histogram of estimated gas density of the starless cores (solid line).
The dotted line represents the ``prestellar" subset, i.e., the gravitationally 
bound cores.  For comparison, the dashed line represents a lognormal with a 
standard deviation of 1.2 in $\ln n({\rm H}_2)$. The error bars are based on 
Poisson statistics with respect to the number in each bin.}
 \label{fig8}
\end{figure}

While the majority of our starless cores are gravitationally unbound and
therefore not prestellar, they 
nevertheless can provide some important information relevant to star formation.
For example, they may represent density enhancements resulting from the
interstellar turbulence widely considered to play a dominant role in the 
determination of the IMF (see, for example, \citealt{pad02,hen08}).  
The probability distribution
function (PDF) of the gas density produced by supersonic
turbulent flow of isothermal gas is well approximated by a lognormal 
form (\citet{elm04} and references therein). It is therefore of interest
to plot a histogram of estimated density values for the L1495 starless cores, 
and this histogram is shown in Fig. \ref{fig8}. These densities are mean values,
calculated by dividing the mass of each core by its total volume, based on
the estimated outer radius, taken to be the deconvolved FWHM of the source.
These densities also correspond to the $n({\rm H}_2)$ values listed in 
Table B2 of Appendix B.  It is evident that, in sharp
contrast to the CMF, the volume density distribution is accurately lognormal 
except for a tail at high densities. A maximum likelihood fit for 
$3\times10^2 \le n({\rm H}_2)[{\rm cm}^{-2}] \le 3\times10^4$ gives
a reduced chi squared value of 0.95, indicating complete consistency with
a lognormal function.

In deriving Fig. \ref{fig8} it was necessary to make allowance for
incompleteness since low-mass cores ($M<0.01\,M_\odot$)
make a significant contribution to the peak of the histogram.
We therefore applied the same incompleteness correction 
as for the starless CMF in Fig. \ref{fig5}. 
In order to investigate the sensitivity to this correction, we have also derived
an alternate version based on a low-mass extrapolation of the ``complete"
portion of the observed
CMF in Fig. \ref{fig5}, the key assumption being that the true CMF of
unbound starless cores is of power-law form.
An almost identical result was obtained, indicating that the shape of
the estimated volume density histogram is not strongly dependent on the specific
assumptions made regarding incompleteness. 

The variance in log density may provide information on the kinetic energy
injection mechanism, the Mach number, and the magnetic field
\citep{mol2012}. To compare Fig. \ref{fig8} with 
turbulence models, however, it must be borne
in mind that what we have plotted is essentially a type of mass-weighted
PDF, as opposed to the volume-weighted version usually presented 
in simulations. More quantitatively, if we make the simplifying
assumption that the density is uniform within an individual core 
(as would be the case for 
pressure-confined clumps\footnote{We do not, of course, assume that all cores 
have the same density. The distribution of core densities would reflect the 
range of external pressures.} that probably form the bulk of the unbound
core distribution) and assume a mass-radius law of
the form $M\propto R^k$, then the core density histogram of Fig. \ref{fig8}
can be represented as a PDF, $P_C(\rho)$, which is related
to the volume-weighted version, $P_V(\rho)$, by:
\begin{equation} 
P_C(\rho)\propto\rho^{3/(3-k)}P_V(\rho)
\label{eq5}
\end{equation}
where $\rho$ is the density.  Furthermore, the 
assumption of lognormal form for $P_V(\rho)$ means that $P_C(\rho)$
is also a lognormal with the same variance in log density, but with
a higher mean density, $\rho_0$. 
It is possible that our density histogram is somewhat narrower than
the true mass-weighted PDF since it does not include the low density
background component. This is probably not a large effect, however, 
since in transforming the volume-weighted PDFs (as presented, for example,
by \citealt{mol2012}) to their mass-weighted counterparts,
the low-density background 
component is de-emphasised. Despite this, the variance of the
distribution is unchanged by the transformation, as illustrated 
by Fig. 6 of \citet{gins13}. This enables 
Fig. \ref{fig8} to be compared
with previously published simulations and theory.

Using the customary notation $s\equiv\ln(\rho/\rho_0)$,
our observations then yield a standard deviation, $\sigma_s$, of 
$1.2\pm0.1$, which is consistent with
simulations of magnetized (MHD) turbulence obtained by \citet{mol2012}
for Mach numbers, ${\cal M}$, in the range $\sim5$--10.  
In those simulations, however, unmagnetized
supersonic turbulence results in an asymmetric low-density tail to the
distribution which is not evident in Fig. \ref{fig8}. The absence of
such a tail seems to argue
for the moderating influence of a magnetic field.

The above value of $\sigma_s$ may also be compared with the results of a recent
study of the turbulence-driven density distribution in a non-star-forming
cloud comprised of G43.17$+$01 and G43.16$-$0.03 by \citet{gins13}, 
who obtained $1.5 < \sigma_s < 1.9$ and used the result to constrain
the compressive part of the driving force. The importance
of the latter study is that it focused on turbulence at the
pre-star-forming stage of the cloud, essentially unaffected by
gravity.  A more detailed comparison of the PDFs of L1495 
and G43.17$+$01/G43.16$-$0.03 may therefore provide evolutionary
information on the density structures.

The one major deviation from lognormality in Fig. \ref{fig8} is the
presence of the high-density tail in the distribution.  We attribute this
tail to the gravitational collapse of cores at the high density end. This
interpretation
is confirmed by restricting the plot to prestellar cores only, as indicated
by the dotted line in the figure. Such
behaviour is predicted by simulations \citep{krit11,collins12,fed13} which
yield model PDFs similar in appearance to Fig. \ref{fig8}. 

The core density above which gravity becomes dominant, i.e.,
the point, $n_{\rm dev}$, at which the histogram in Fig. \ref{fig8} deviates 
from lognormal behaviour, is $\sim10^5\,{\rm cm}^{-3}$. A similar
situation can be inferred in Aquila \citep{kon15}, whereby the volume 
density PDF shows the tentative presence of a power-law tail above
$\sim10^5\,{\rm cm}^{-3}$. 
The importance of the deviation value is that it can be related directly to 
the background volume
density threshold for star formation, $n_{\rm crit}$, a quantity
which so far has been difficult to obtain observationally \citep{kai14}.
The relationship is based on the fact
that the volume density contrast of a critical B-E sphere over the
background is $\sim2.4$ (cf. \citealt{and07}). We thereby estimate $n_{\rm crit} 
\sim4\times10^4\,{\rm cm}^{-3}$.

The above value may be compared with the expected threshold based on
the collapse of supercritical filaments, as discussed by \citet{and14}.
Specifically, the critical line mass of 16 $M_\odot{\rm pc}^{-1}$,
together with the characteristic filamentary width of 0.1 pc, 
implies a critical number density $n_{{\rm H}_2}\sim2\times10^4\,
{\rm cm}^{-3}$. This is consistent with our $n_{\rm crit}$ estimate,
given the factor of $\sim2$ uncertainty in the latter.
Our estimate is also consistent with the value
$\sim3\times10^4\,{\rm cm}^{-3}$ expected in the theoretical picture
described by \citet{pad14}, but an order of magnitude
larger than an observational value, $(5\pm2)\times10^3\,{\rm cm}^{-3}$, 
obtained by \citet{kai14} for other nearby regions via a novel technique 
involving column density PDFs.

The average logarithmic slope of the high density tail of the
histogram in Fig. \ref{fig8}
is $\sim -1$, but with considerable uncertainty (values
in the range $-0.8$ to $-1.6$ are consistent with the data).
Taking these values together with the power-law index, 
$k=1$, appropriate to critical B-E spheres, 
Eq. (\ref{eq5}) would imply that the slope of the 
high density tail of the volume-weighted version, $P_V(\rho)$, should be 
in the range $\sim -2.3$ to $-3.1$.  These
slopes are somewhat steeper than the theoretical asymptotic value of $-1.54$
\citep{giri14} and the values suggested by simulations, which fall
in the range $[-7/4,\,-3/2]$ \citep{krit11}. Further progress in comparing the 
observational characteristics of prestellar cores with theoretical prediction 
must await a more detailed analysis. Indeed, a fruitful avenue may be
to apply our observational core detection procedure directly to the simulated 
images produced by the gravo-turbulence models.

\begin{figure}
\includegraphics[width=88mm]{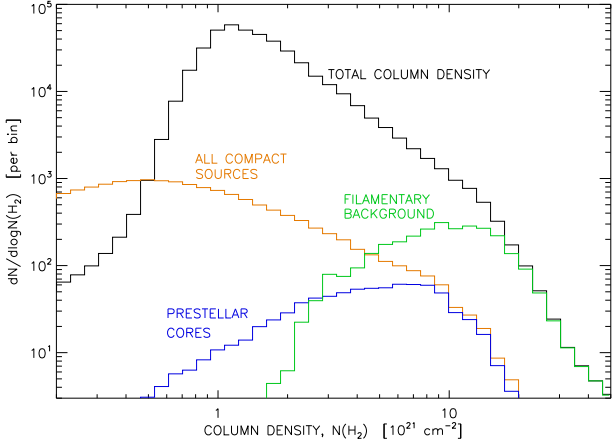}
 \caption{Column density PDFs. The black histogram represents the distribution 
of total column density values averaged over an effective resolution
element of FWHM $18.2''$, and is truncated
on the left at the approximate level of the noise. The gold histogram 
represents the distribution after the subtraction of filamentary background, 
i.e., it represents the net contribution from
compact ($\stackrel{<}{_\sim}4'$) sources in the field. The blue
histogram represents the contribution from prestellar cores only.
The green histogram represents the distribution of total column density in 
the immediate vicinity (within two core radii) of prestellar cores, and
shows that the highest column densities are dominated by the filamentary
backgrounds in those neighbourhoods.}
 \label{fig9}
\end{figure}

Previous analyses of the observationally determined density structure of
star-forming regions have focused on PDFs of column density rather than volume
density. Those PDFs have been characterised by a lognormal at low
column densities which becomes dominated by a power law tail at the high end
\citep{kain09,sch13,sch15,ben15}. The presence of the lognormal
component has, however, been questioned by \citet{lom15} who argue
that column density PDFs of molecular clouds can be accounted for entirely
by truncated power laws. Conversely, the reality of some of the power-law
components has been questioned by \citet{brunt15} who has shown that
spurious power-law tails can arise if proper account is not taken of
noise and warm background components. The PDF of total column density in L1495,
shown by the black histogram in Fig. \ref{fig9} does appear to have
distinct power-law behaviour for all but the lowest column densities.
To interpret the PDF in light of the various possible effects, it is helfpul
to separate the contributions of starless
cores and filamentary structure, by replotting the 
PDF after subtraction of the filamentary background (the {\it getsources\/}
cleaned background image).
The resulting histogram is shown in gold in Fig. 
\ref{fig9}. It represents the column density PDF 
associated with compact sources on size scales $\stackrel{<}{_\sim}4'$ and
is seen to be qualitatively similar to the PDF of volume density in 
Fig. \ref{fig8}. In particular, the high density tail in both cases is
fully accounted for by the prestellar cores which have been identified via 
source extraction and whose contribution is shown
by the blue histogram in Fig. \ref{fig9}. These tails, therefore, cannot be 
attributed to artifacts of the type discussed by \citet{brunt15}. 

With regard to the histogram of total column density, Fig. \ref{fig9} shows 
that the cores themselves do not account for the highest column densities
in the region---the filamentary background is important also, as
illustrated by the green histogram in that figure.  Although
the peak column densities occur at the locations of prestellar cores,
the filamentary contribution is dominant at those locations, a situation
similar to that in Aquila \citep{kon15}.

\section[]{Conclusions}
The principal conclusions from this study can be summarized as follows:
\begin{enumerate}
  \item We have obtained a census of prestellar cores in L1495 (total = 52) 
which is $\sim85$ \% complete for $M/M_\odot>0.1$;
  \item The overall starless CMF is not well represented by a lognormal
function and shows no evidence of a low-mass turnover. Instead, it 
is consistent with power-law behaviour similar to that shown by
clumps seen in various isotopologues of CO;
  \item The prestellar CMF of L1495 resembles
the stellar system IMF as has been found previously for other 
star-forming regions, albeit with rather low number statistics;
  \item All of the inferred prestellar cores lie on filamentary structure and 
the associated background column density distribution is sharply peaked near 
the critical value for cylindrical stability against gravitational collapse;
  \item The PDF of volume density for starless cores is consistent with the
effect of self gravity on magnetized supersonic turbulence. Indeed, it is
characterised by a lognormal except for the presence of a high density tail 
attributable entirely to collapsing prestellar cores. Note that this 
conclusion applies to
the starless core population only and does not include the contribution
of the filamentary background.
\end{enumerate}

\section*{Acknowledgments}

We thank the referee for helpful comments.
SPIRE has been developed by a consortium of institutes led by
Cardiff Univ. (UK) and including: Univ. Lethbridge (Canada);
NAOC (China); CEA, LAM (France); IFSI, Univ. Padua (Italy);
IAC (Spain); Stockholm Observatory (Sweden); Imperial College
London, RAL, UCL-MSSL, UKATC, Univ. Sussex (UK); and
Caltech, JPL, NHSC, Univ. Colorado (USA). This development
has been supported by national funding agencies: CSA (Canada);
NAOC (China); CEA, CNES, CNRS (France); ASI (Italy);MCINN
(Spain); SNSB (Sweden); STFC, UKSA (UK); and NASA (USA).
PACS has been developed by a consortium of institutes led by
MPE (Germany) and including UVIE (Austria); KU Leuven, CSL,
IMEC (Belgium); CEA, LAM (France); MPIA (Germany); INAFIFSI/
OAA/OAP/OAT, LENS, SISSA (Italy); IAC (Spain). This development
has been supported by the funding agencies BMVIT
(Austria), ESA-PRODEX (Belgium), CEA/CNES (France), DLR
(Germany), ASI/INAF (Italy), and CICYT/MCYT (Spain).
This work has received support from the European Research Council under the 
European Union's Seventh Framework Programme (ERC Advanced Grant Agreement 
no. 291294, `ORISTARS') and from the French National Research Agency 
(Grant no. ANR-11-BS56-0010 -- `STARFICH').
This research has made use of the SIMBAD database, operated at CDS,
Strasbourg, France.
It has also made use of the NASA/IPAC Extragalactic Database (NED) 
which is operated by the Jet Propulsion Laboratory, California Institute 
of Technology, under contract with the National Aeronautics and Space 
Administration.

\appendix

\section[]{Estimation of catalogue completeness}

The completeness of starless core detection may be estimated by repeating the 
full source extraction and source classification procedure using simulated
data for a set of model cores whose spatial distribution, spatial density,
and physical parameters resemble those in the
actual L1495 cloud. In doing so, one must consider that 
the detectability of a core is not fully characterised by its mass. The mean 
dust temperature is another important factor and that quantity is, in turn, 
determined by other physical parameters whose distributions can lead to 
different possible flux densities for a given mass. In particular,
the cooler and more highly condensed prestellar cores have different
detectability characteristics from those of the general unbound starless
population which is more diffuse. We have
therefore modelled the completeness of those two object types separately.

\begin{figure}
\hspace*{-0.2cm}\includegraphics[width=90mm]{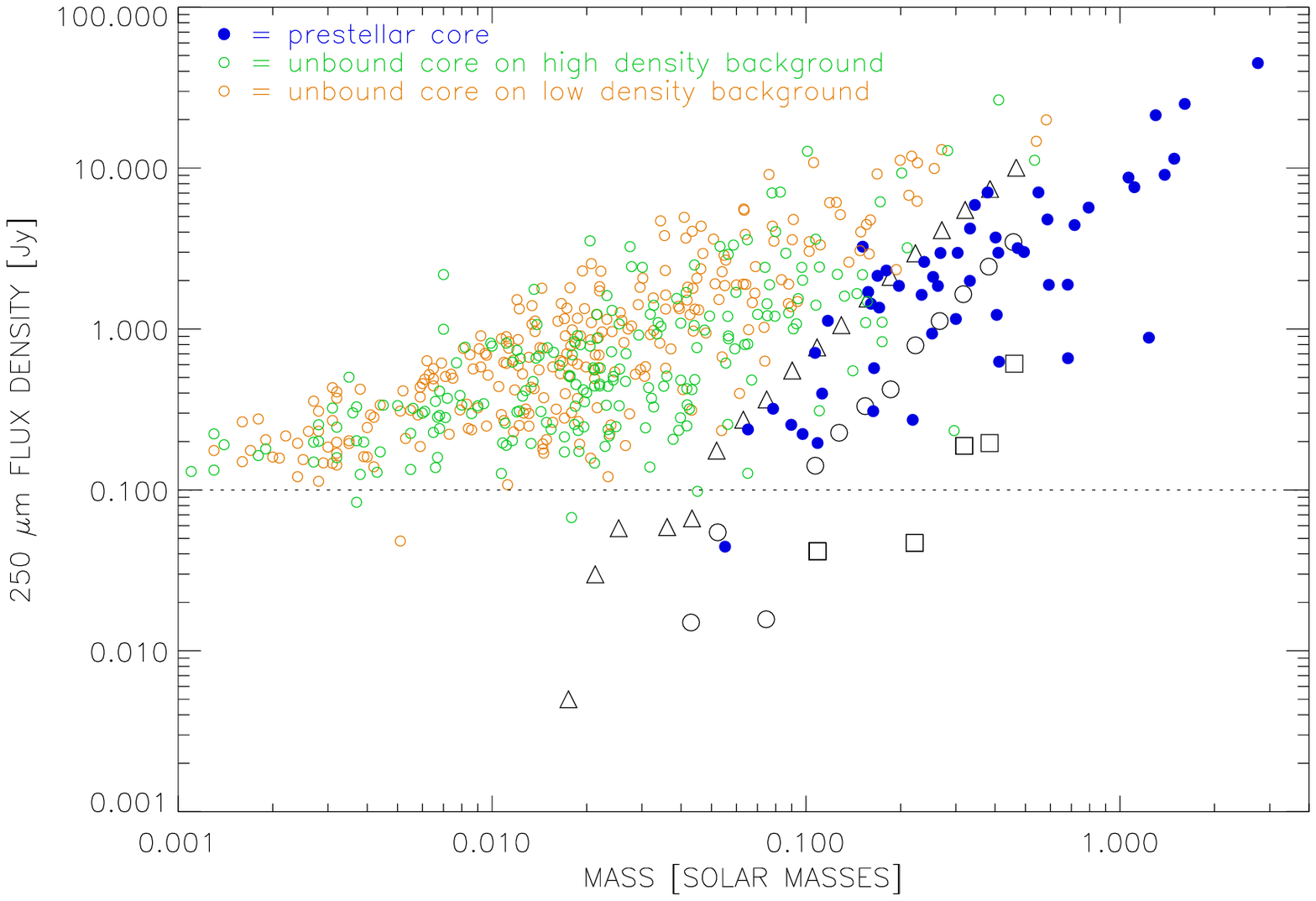}
 \caption{The 250 $\mu$m flux densities of observed and model cores as a 
function of mass.  Black symbols represent the model prestellar cores 
used in the completeness simulation, taken to be critical B-E spheres.
They are plotted for three density regimes, indicated by triangles 
(low density case), open circles (moderate density case) and squares (high 
density case). Also shown for comparison are the 
observational values for the extracted sources, represented by the coloured
symbols. These are divided into
prestellar cores (blue filled circles), and gravitationally unbound
cores (open circles, where green and gold designate those seen against high
and low column density backgrounds, respectively, based on a threshold
of $A_V=7$ mag).  For reference, the horizontal dotted
line represents the approximate flux density sensitivity limit.}
 \label{figA1}
\end{figure}

Following \citet{kon15}, we have modelled the bound (prestellar) cores as 
critical B-E spheres. The radial temperature distribution was 
determined from a dust radiative transfer model (Men'shchikov et al., in 
preparation). Three populations of bound cores were generated, characterised
by median central density values of
approximately $4\times10^6$ cm$^{-3}$ (low density set), 
$2\times10^7$ cm$^{-3}$ (moderate density set), 
and $6\times10^7$ cm$^{-3}$ (high density set), with a wide range 
(4--5 orders of magnitude) within each set. In all cases, the
cores were embedded in a cloud with $A_V =3.7$ mag.
The behaviour of 250 $\mu$m flux density as a function of mass for the three
sets is shown in Fig. \ref{figA1} and may be 
compared with the observed
values of extracted prestellar cores.

With regard to the unbound starless cores,
Fig. \ref{figA1} shows that these objects are detected
at masses considerably below those of prestellar cores, both in regions
of high and low column density.  Many of these objects are probably
transient structures which are not only gravitationally unbound but may
also not even be pressure confined.  Their importance to star formation
studies lies in the fact that, in terms of mass distribution, they appear to 
form a continuous sequence that includes prestellar cores. The latter objects 
represent the upper end of the mass range, defined by the point at
which self-gravity becomes dominant. The unbound cores  may therefore provide
useful constraints on dynamical models of the interstellar turbulence that
gives rise to the prestellar subset. 

At the low end of the mass range ($M\stackrel{<}{_\sim}0.1\,M_\odot$),
all of the detected objects are either unresolved
or marginally resolved and hence for a given mass and background rms,
the only quantity which determines detectability is then the dust temperature.
As discussed earlier, Fig. \ref{fig4} suggests a negative correlation
between mass and temperature although, given the detection limits,
it is not clear that the correlation persists down to the lowest masses.
However, if we assume that the lowest-mass condensations are diffuse,
then their dust temperatures can be expected to be systematically high
due to easier penetration by the ISRF, and hence
be more detectable than a prestellar core of the same mass.
For larger masses and hence larger radii, however, cores become spatially 
resolved and detectability is then
determined by the peak contrast, $N_{{\rm H}_2}^{\rm pk}/N_{\rm bg}$,
of column density with respect to the local background. In that regime,
a prestellar core becomes more detectable than an unbound core of the same
mass.

It is, of course, possible that the detected objects are not representative of 
the entire population of unbound starless cores. For example, the detection
limit in Fig. \ref{fig4} leaves open the possibility of a large population of
cool objects of low mass. Likewise, we cannot rule
out the existence of a substantial population of objects of very low density
contrast, at all sizes, which are not detected by {\it getsources\/}.
In order to provide a meaningful definition of completeness for unbound
starless cores, we therefore need a more precise observational definition.
So as a working set of criteria, we take such objects to be gravitationally 
unbound condensations whose sizes are approximately consistent with the
\citet{lars81} mass-radius relation and whose dust temperatures are consistent
with the mass-temperature behaviour shown in Fig. \ref{fig4}.
We have made an assessment of the detection completeness of such objects
using simulations based on empirically determined values of physical parameters.
The latter were based on the following functional forms,
adapted from \citet{marsh2014}, for 
the radial dependence of molecular hydrogen number density, $n(r)$, and dust 
temperature, $T(r)$, assuming spherical symmetry:
\begin{eqnarray}
    n(r) & = & n_0/[1 + (r/r_0)^2]  \label{eq2} \\
    T(r) & = & T_0 + (T_{\rm out}\!-\!T_0)r/r_{\rm out}
    \label{eq3}
\end{eqnarray}
where $r_0$ represents the radius of an inner
plateau, $r_{\rm out}$ is the outer radius of the core, $T_0$ is the central 
core temperature, and $T_{\rm out}$ is the temperature at the outer radius.
In relating $n(r)$ to the corresponding profile of mass 
density, we assumed a mean molecular weight, $\mu$, of 2.8 \citep{roy2014}. 
The assumed mass distribution was based on the \citet{kroupa01} IMF,
and the slope of the mass-radius relation was taken as 1.9
\citep{lars81}. The relative values
of $r_0$ and $r_{\rm out}$ and their statistical dispersions were based 
on the results of \citet{marsh2014}. The temperatures
$T_0$ and $T_{\rm out}$ were also based on the latter results, with
a constant value of 11 K assumed for $T_{\rm out}$.
The emergent intensity was then calculated using the 
opacity law given by Eq. (\ref{eq1}).  

\begin{figure}
\includegraphics[width=84mm]{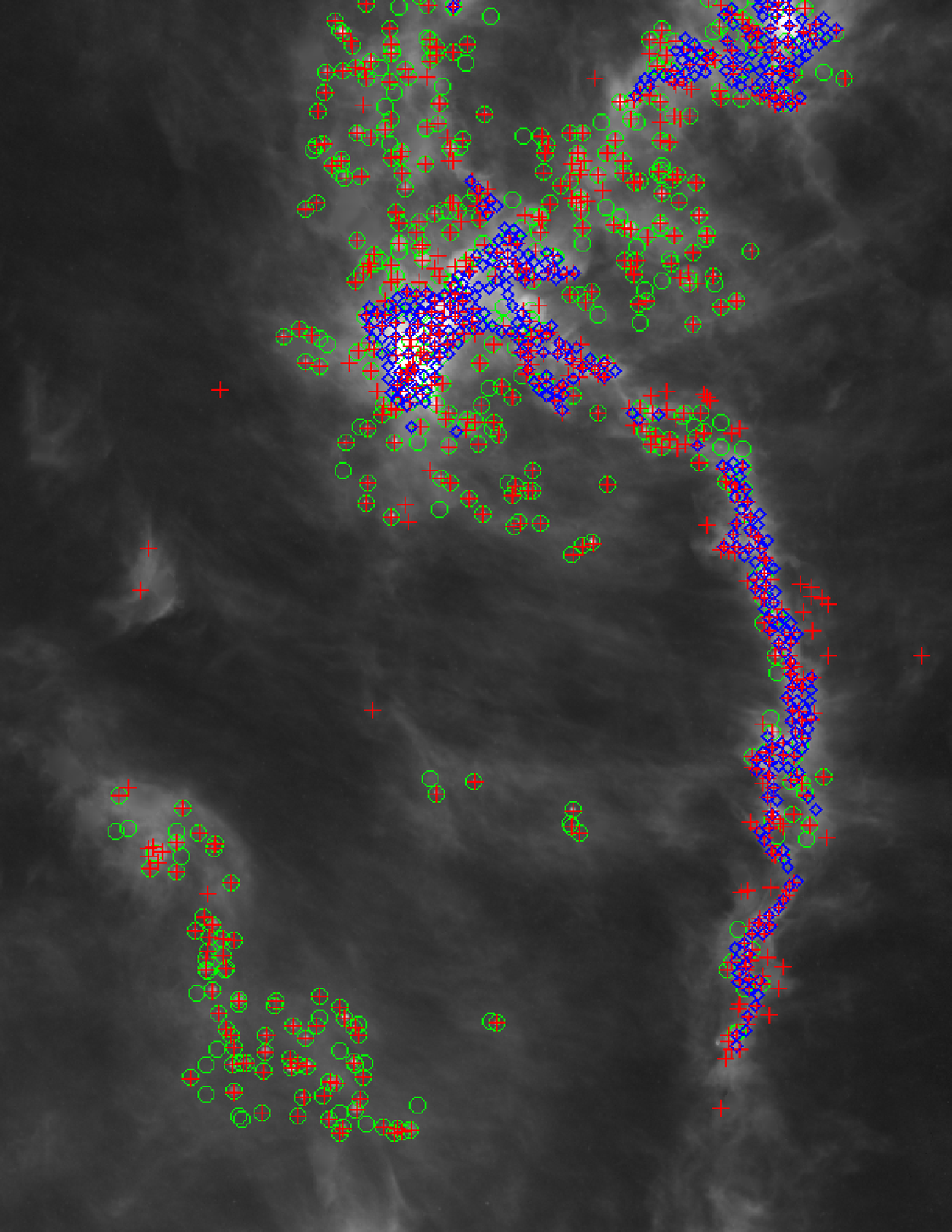}
 \caption{Simulated SPIRE 250 $\mu$m image, used for evaluation of the
catalogue completeness. The field dimensions and intensity scale are the same 
us for Fig.~\ref{fig1}.  Green circles represent empirically modelled
cores which may be interpreted as the dust-emission counterparts of CO clumps,
blue diamonds represent bound (prestellar) model cores, and red crosses
represent core candidates extracted from the simulated data.}
 \label{figA2}
\end{figure}

Using the above source models, we have constructed synthetic images for 384 
critical B-E cores and 600 additional objects whose parameter values 
were based on empirical data. The latter subset was designed to represent 
the general unbound starless population. 
Each model image was convolved
with the {\it Herschel\/} beam at the corresponding wavelength using
published beam data. Specifically,
the PACS beams were obtained by azimuthal averaging the 
observationally determined PSFs from \citet{lutz2012}. For SPIRE,
azimuthally symmetric PSFs were obtained using the radial profiles
from \citet{griffin13}.
The model images were then sprinkled onto noisy backgrounds corresponding to
the {\it getsources\/} cleaned background images \citep{men12} at the five
wavelengths.  They were distributed randomly over the portions of the image
dominated by dense filamentary material which, for the empirically-modelled
objects, were taken as the regions in which the 250 $\mu$m background intensity 
exceeded 57 MJy sr$^{-1}$ after applying the Planck offset correction.
This intensity level corresponds to $A_V\stackrel{<}{_\sim}5$ mag over
most of the region.  The B-E population, on the other hand,
was distributed over the regions for which the local column density 
exceeded $5\times 10^{21}$ cm$^{-2}$ ($A_V > 5$ mag).
Fig. \ref{figA2} shows the simulated 250 $\mu$m image together with the
locations of the model sources.

\begin{figure}
\hspace*{-0.4cm}\includegraphics[width=90mm]{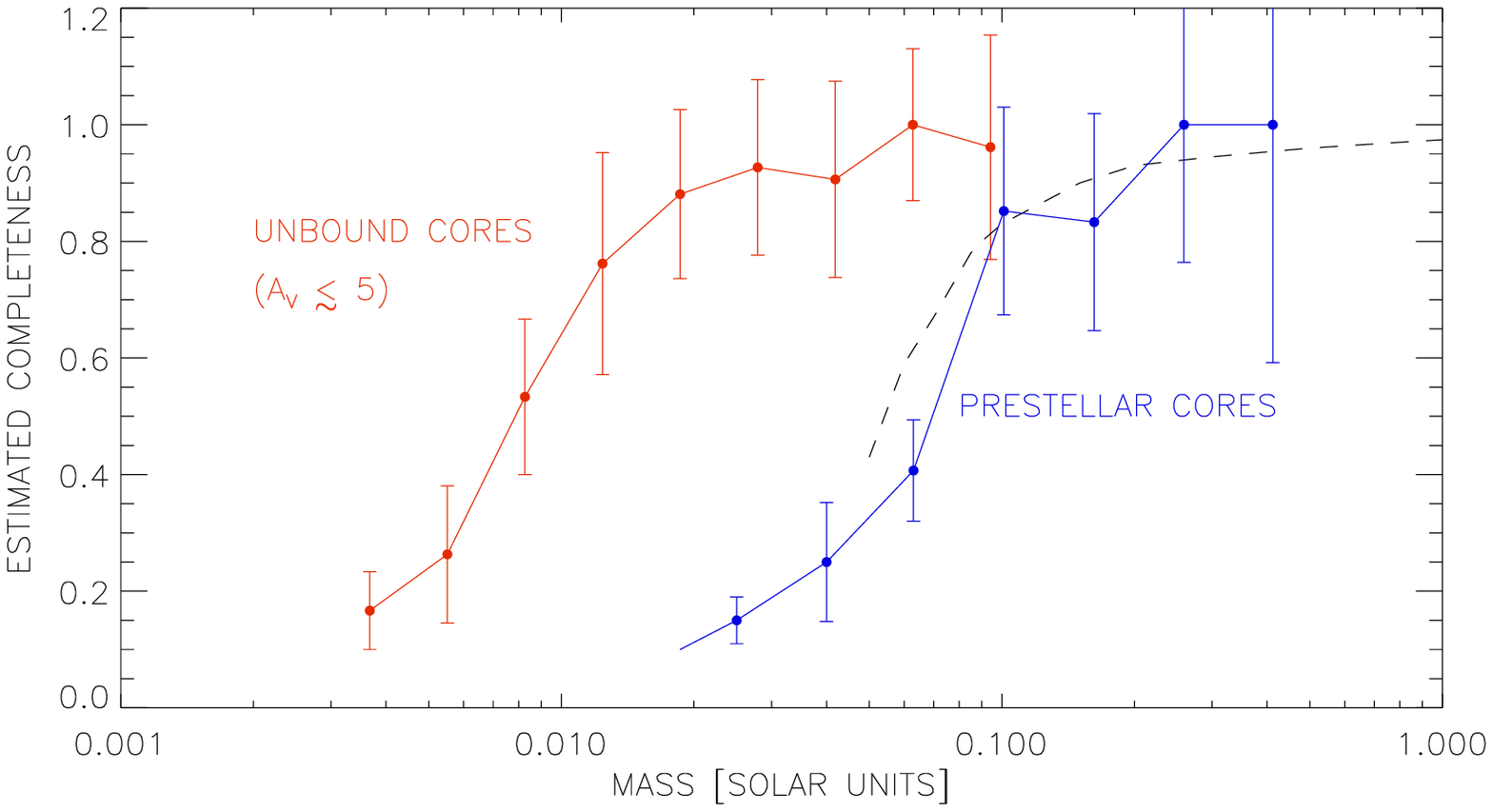}
 \caption{Catalogue completeness as a function of mass based on
simulated data for 
unbound cores and bound (prestellar) cores, 
represented by the red and blue curves, respectively.
Error bars represent Poisson statistics only.
The curve for unbound cores is representative of moderate 
($A_V\stackrel{<}{_\sim}5$ mag) backgrounds.
In that regime, low mass unbound cores ($M/M_\odot<0.1$) are more easily
detectable than their prestellar counterparts due to their systematically
higher temperatures. The simulated completeness for prestellar cores may be
compared with the expected behaviour based on the model described
by \citet{kon15}, indicated by the black dashed curve.}
 \label{figA3}
\end{figure}

Having thus generated synthetic images at the five wavelengths
for the same $4^\circ\times2^\circ$ field as the {\it Herschel\/}
observations, the same source detection and classification procedure,
as discussed in Section 2, was used to produce a set of core 
candidates---their
locations are also indicated in Fig. \ref{figA2}.
Comparison of recovered sources with the ``truth" table from the
simulations enabled completeness curves to be derived. These curves
were derived separately for the unbound and bound cores
and the results are shown in Fig. \ref{figA3}. 
In the case of unbound cores, the completeness curve is representative
of moderate backgrounds for which 
$A_V\stackrel{<}{_\sim}5$ mag. The error bars 
represent Poisson errors only and do not incorporate the additional effects 
of model errors which are difficult to quantify. But subject to that
caveat, we estimate that our census of unbound cores is $\sim85$\%
complete for $M/M_\odot>0.015$ at $A_V\stackrel{<}{_\sim}5$ mag.

For the prestellar cores
we have compared the results of our numerical simulation with the 
predicted completeness obtained from the model described in Appendix B.2 of
\citet{kon15}, which is based on
critical B-E spheres. 
The predicted curve for L1495 is indicated by the black dashed curve 
in Fig. \ref{figA3} and is seen to be quite consistent with the simulation.
Since this curve is model-based it represents {\it true\/} mass as opposed to
observationally derived mass, the latter of which may be systematically 
underestimated by up to a factor of 2 as discussed in Section 6.
Based on these results we estimate that the census of bound cores is
$\sim85$\% complete for $M/M_\odot>0.1$.

\section[]{The catalogue and derived core properties}

Tables B1 and B2 show example portions of the dense core catalogue and the 
derived properties for dense cores, respectively. The complete
tables are available in the on-line version of the paper\footnote{The
tables are also available, together with the full set of HGBS data products,
from http://gouldbelt-herschel.cea.fr/archives}. 
The quoted values
of integrated flux for non-detections are, in some cases, negative, 
particularly at 70 $\mu$m wavelength. Such values,
in conjunction with their standard deviations,
convey statistical information necessary for optimal fitting of SEDs using
the maximum likelihood criterion. They can also be physically meaningful
since, e.g., starless cores can be seen in absorption against a warmer
background at 70 $\mu$m. 

In addition to starless cores, both tables
contain entries for dense cores in which protostars are embedded. For 
present purposes, an embedded core is defined as one in which a significant
compact source of 70 $\mu$m emission is present within its 50\% column density 
contour. Their derived properties
are more uncertain and their masses may be underestimated by a factor
of $\sim2$. We detect 11 such objects in L1495. They are not 
considered further in the present work, but
will be discussed in a forthcoming paper.

The rightmost column of Table B1 indicates matches with objects listed in the
SIMBAD database and includes several previous surveys of dense cores.
In most cases, those surveys were made at lower resolution
and typically have blended the core either with neighbouring cores or with 
associated background
filamentary structure.  One example is LDN 1495N, whose mass and radius are 
listed as 1.85 $M_\odot$ and 0.07 pc, respectively, in the NH$_3$ survey of
\citet{bens89}, whereas the corresponding {\it Herschel\/} values are
1.39 $M_\odot$ and 0.050 pc, respectively.  More extreme examples are to
be found in comparisons with the C$^{18}$O catalogue of \citet{onishi1996},
in which objects 2, 3, and 11 have masses in the range 16.6--46.0 $M_\odot$,
considerably in excess of their {\it Herschel\/} counterparts. 
A closer correspondence is found between our {\it Herschel\/} cores
and the H$^{13}$CO$^+$ cores of \citet{onishi2002}, as discussed
by \citet{marsh2014}.

Comments in the far right column of Table B2 indicate 
failed SED fits (see Section 4). They also indicate cases in which the
source was spatially unresolved in column density. In those cases,
the mean peak column densities and volume densities were derived using
the observed source size only. 

\begin{landscape}
\begin{table}
 \begin{minipage}{200mm}
  \caption{Catalogue of dense cores in L1495. The full table is presented in the
on-line version of the paper. A portion is reproduced here to show
form and content.}
  \tiny
  \begin{tabular}{@{}cccccccccccp{0.6cm}p{0.6cm}cc@{}}
  \hline
   Src\#\footnote{Source number in catalogue.} &
   Source name\footnote{Source name. The full name consists of the prefix ``HGBS\_J" followed directly by the tabulated label created from J2000 sexagesimal coordinates.}  & 
      RA$_{2000}$  & 
     Dec$_{2000}$  & 
(Sig$_{70}$)\footnote{Detection significance from monochromatic single 
scales in the image at the specified wavelength. Non-detections are labelled as 0.0.} &
  ($I_{70}^{\rm pk}$)\footnote{Estimate of the peak intensity.} &
  $\sigma(I_{70}^{\rm pk})$\footnote{Uncertainty in peak intensity.} &
  ($I_{70}^{\rm pk}/I_{\rm bg}$)\footnote{Contrast of the peak intensity 
over the local background.} &
  ($I_{070}^{\rm conv,500}$)\footnote{Peak intensity after smoothing to a $36''\!\!\!.\,3$ beam.} &
  ($S_{70}$)\footnote{Estimate of the total flux and its uncertainty.} &
  $\sigma(S_{70})$\footnote{Uncertainty in total flux.} &
  ($a_{70}$)\footnote{Estimate of source size (FWHM) along major axis.} &
  ($b_{70}$)\footnote{Estimate of source size (FWHM) along minor axis.} &
  (PA$_{70})$\footnote{Position angle of source major axis, east of north.} &
Cont'd \\
 & HGBS\_J$^*$ & [h m s] & [$^\circ\,'\,''$] & & [Jy/beam] & [Jy/beam] &  & [Jy/beam$_{500}$] & [Jy] & [Jy] & [$''$] &[$''$] & [$^\circ$] & $\rightarrow$ \\
 \hline
     1 & 040924.1+284723 & 04 09 24.10 &   +28 47 23 &    0.9 &  -3.81e-03 & 4.3e-03 &    0.00 &  -5.55e-02 &   5.41e-04 & 2.2e-02 &    85 &   44 &   77 \\
    17 & 041042.8+281925 & 04 10 42.84 &   +28 19 25 & 0.0 &   2.92e-02 & 1.2e-02 &    0.00 &   2.29e-01 &   8.02e-02 & 3.4e-02 &    29 &   27 &   50 \\
    76 & 041335.4+282117 & 04 13 35.45 &   +28 21 17 &    2.4 &  -1.59e-02 & 6.7e-03 &    0.00 &  -3.00e-01 &   4.13e-01 & 7.8e-02 &   143 &  116 &  153 \\
   421 & 041840.7+281916 & 04 18 40.67 &   +28 19 16 & 2245.0 &   2.94e+01 & 5.5e-02 &   38.12 &   3.49e+01 &   5.16e+01 & 1.1e-01 &     9 &    8 &  159 \\
\hline \\
\end{tabular}
  \begin{tabular}{@{}cccp{0.8cm}p{1.0cm}ccp{0.2cm}p{0.2cm}p{0.4cm}cccp{0.8cm}p{1.0cm}ccp{0.2cm}p{0.2cm}p{0.6cm}p{0.8cm}@{}}
  \hline
Sig$_{160}$ & 
$I_{160}^{\rm pk}$  &
$\sigma(I_{160}^{\rm pk})$  &
$I_{160}^{\rm pk}/I_{\rm bg}$ & 
$I_{160}^{\rm conv,500}$ &
$S_{160}$ &
$\sigma(S_{160})$ &
$a_{160}$ & 
$b_{160}$ & 
PA$_{160}$ &
Sig$_{250}$ & 
$I_{250}^{\rm pk}$ & 
$\sigma(I_{250}^{\rm pk})$ & 
$I_{250}^{\rm pk}/I_{\rm bg}$ & 
$I_{250}^{\rm conv,500}$ &
$S_{250}$ &
$\sigma(S_{250})$ &
$a_{250}$ & 
$b_{250}$ & 
PA$_{250}$ & Cont'd \\
  & [Jy/beam] & [Jy/beam] &  & [Jy/beam$_{500}$] & [Jy] & [Jy] & [$''$] &[$''$] & [$^\circ$] & 
  & [Jy/beam] & [Jy/beam] & & [Jy/beam$_{500}$] & [Jy] & [Jy] & [$''$] &[$''$] & [$^\circ$] & $\rightarrow$ \\
 \hline
      6.5 &   1.04e-01 & 2.6e-02 &    0.56 &   5.04e-01 &   1.02e+00 & 8.1e-02 &    56 &   35 &     56 &     8.8 &   2.13e-01 & 3.5e-02 &    1.07 &    0.483 &   1.34e+00 & 8.1e-02 &     44 &   34 &  132 \\
   0.0 &   1.87e-02 & 2.2e-02 &    0.00 &   3.47e-02 &   4.33e-02 & 3.8e-02 &    20 &   17 &     69 &     8.1 &   1.97e-01 & 3.0e-02 &    0.46 &    0.330 &   3.98e-01 & 3.1e-02 &     32 &   24 &   82 \\
      6.1 &   9.87e-03 & 2.3e-02 &    0.00 &   6.44e-02 &   1.62e+00 & 1.1e-01 &   176 &   53 &     86 &    18.8 &   6.37e-01 & 5.4e-02 &    1.47 &    1.996 &   1.14e+01 & 2.7e-01 &     72 &   63 &   27 \\
    927.9 &   2.13e+01 & 4.5e-01 &    6.91 &   2.32e+01 &   2.06e+01 & 5.3e-01 &    14 &   14 &     90 &   670.8 &   1.06e+01 & 5.5e-01 &    3.11 &   10.174 &   7.08e+00 & 4.8e-01 &     18 &   18 &  128 \\
\hline \\
\end{tabular}
  \begin{tabular}{@{}cccp{0.8cm}p{1.0cm}ccp{0.2cm}p{0.2cm}p{0.4cm}cccp{0.8cm}ccp{0.2cm}p{0.2cm}p{0.6cm}p{0.8cm}@{}}
\hline
Sig$_{350}$ & 
$I_{350}^{\rm pk}$ & 
$\sigma(I_{350}^{\rm pk})$ & 
$I_{350}^{\rm pk}/I_{\rm bg}$ & 
$I_{350}^{\rm conv,500}$ &
$S_{350}$ &
$\sigma(S_{350})$ &
$a_{350}$ & 
$b_{350}$ & 
PA$_{350}$ &
Sig$_{500}$ & 
$I_{500}^{\rm pk}$ & 
$\sigma(I_{500}^{\rm pk})$ & 
$I_{500}^{\rm pk}/I_{\rm bg}$ & 
$S_{500}$ &
$\sigma(S_{500})$ &
$a_{500}$ & 
$b_{500}$ & 
PA$_{500}$ & Cont'd \\
 & [Jy/beam] & [Jy/beam] &  & [Jy/beam$_{500}$] & [Jy] & [Jy] & [$''$] &[$''$] & [$^\circ$] &
 & [Jy/beam] & [Jy/beam] &  &  [Jy] & [Jy] & [$''$] &[$''$] & [$^\circ$] & $\rightarrow$ \\
 \hline
     12.7 &   2.72e-01 & 6.9e-02 &    1.11 &   4.25e-01 &   1.17e+00 & 1.2e-01 &     50 &   41 &   134 &     13.8 &   2.34e-01 & 5.8e-02 &    0.67 &   5.11e-01 & 6.8e-02 &    57 &   43 &    135 \\
     11.6 &   2.88e-01 & 5.3e-02 &    0.60 &   3.31e-01 &   4.05e-01 & 4.8e-02 &     34 &   25 &    68 &     14.0 &   2.20e-01 & 7.3e-02 &    0.50 &   3.23e-01 & 6.7e-02 &    55 &   37 &     80 \\
     32.3 &   9.26e-01 & 7.6e-02 &    1.41 &   1.73e+00 &   1.06e+01 & 2.8e-01 &     81 &   68 &   168 &     41.4 &   9.52e-01 & 7.4e-02 &    0.76 &   5.24e+00 & 1.6e-01 &   103 &   61 &    152 \\
    348.6 &   5.24e+00 & 7.2e-01 &    1.83 &   5.21e+00 &   4.40e+00 & 6.6e-01 &     25 &   25 &   147 &    157.5 &   2.36e+00 & 8.2e-01 &    0.96 &   1.88e+00 & 7.5e-01 &    36 &   36 &    158 \\
 \hline \\
\end{tabular}

  \begin{tabular}{@{}ccccccccccccc@{}}
  \hline
Sig($N_{{\rm H}_2}$)\footnote{Detection significance from single scales on 
high-resolution column density map.} & 

($N_{{\rm H}_2}^{\rm pk}$)\footnote{Peak column density.} &

($N_{{\rm H}_2}^{\rm pk}/N_{\rm bg}$)\footnote{Contrast of peak column density
over the local background.} &

($N_{{\rm H}_2}^{\rm conv,500}$)\footnote{Peak column density after smoothing to a $36''\!\!\!.\,3$ beam.} &

($N_{{\rm H}_2}^{\rm bg}$)\footnote{Local background column density.} &

($a[N_{{\rm H}_2}$])\footnote{FWHM along major axis in column density.} &

($b[N_{{\rm H}_2}$])\footnote{FWHM along minor axis in column density.} &

(${\rm PA}[N_{{\rm H}_2}$])\footnote{Position angle of major axis in column density.} &

$N_{\rm SED}$\footnote{Number of {\it Herschel\/} bands in which the source is significant (Sig$_\lambda>5$) and has a positive flux density, excluding
the column density plane.} &

CSAR\footnote{Equal to 1 if source found independently by CSAR, and 0 otherwise.} & 

Core type\footnote{1=starless(unbound); 2=prestellar; 3=candidate prestellar; 4=dense core with embedded protostar.} &

SIMBAD\footnote{SIMBAD association:
ID of nearest match if within $1'$ of {\it Herschel\/} position. A key to
the abbreviations used is given in the online version.} &

Comments \\

  & [$10^{21}\,{\rm cm}^{-2}$] &  & [$10^{21}\,{\rm cm}^{-2}$] & [$10^{21}\,{\rm cm}^{-2}$]& [$''$] &[$''$] & [$^\circ$] & & &  & \\
 \hline
    8.4 &    0.8 &    0.25 &    2.2 &   11.4 &   56 &   35 &  119 &     4 &     1 &     1 &   \\
   13.2 &    1.7 &    0.62 &    2.8 &   16.1 &   33 &   18 &   41 &     4 &     0 &     3 &   \\
   62.3 &    6.7 &    0.93 &   23.6 &   33.4 &   81 &   58 &  145 &     4 &     1 &     2 & [OMK2002] 3 \\
   73.9 &    7.5 &    1.04 &    9.3 &   28.2 &   22 &   18 &  157 &     5 &     1 &     4 & V892 Tau \\
 \hline
\end{tabular}
\end{minipage}
\end{table}
\end{landscape}

\begin{landscape}
\begin{table}
 \begin{minipage}{200mm}
  \caption{Derived properties of dense cores in L1495. The full table is presented in the
on-line version of the paper. A portion is reproduced here to show
form and content.}
  \tiny
  \begin{tabular}{@{}p{0.2cm}cccp{0.6cm}p{0.8cm}p{0.4cm}p{0.8cm}p{0.4cm}p{0.8cm}p{1.0cm}p{1.0cm}p{1.0cm}p{1.0cm}cp{0.8cm}p{0.6cm}p{0.3cm}c@{}}
  \hline
   Src\#\footnote{Source number in catalogue} &
   Source name\footnote{Source name. The full name consists of the prefix ``HGBS\_J" followed directly by the tabulated label created from J2000 sexagesimal coordinates.}  & 
      RA$_{2000}$  & 
     Dec$_{2000}$  & 
($R_{\rm c}^{\rm obs}$)\footnote{Core radius as observed, derived from the high-resolution ($18''\!\!.2$) column density map, assuming 140 pc distance.} &
($R_{\rm c}^{\rm dec}$)\footnote{Core radius, beam-deconvolved; derived from the high-resolution ($18''\!\!.2$) column density
map, assuming 140 pc distance.} &
($M_{\rm c}$)\footnote{Estimated core mass.} &
$(\sigma\! M_{\rm c})$\footnote{Uncertainty in core mass.} &
($T_{\rm d}$)\footnote{Estimated dust temperature.} &
$(\sigma\!T_{\rm d})$\footnote{Uncertainty in dust temperature.} &
($N_{H_2}^{\rm pk}$)\footnote{Peak H$_2$ column  density at the resolution of the
500 $\mu$m data, derived from a modified blackbody SED fit to the core peak flux densities
measured in a common $36''\!\!.3$ beam at all wavelengths.} &
($\bar{N}_{H_2}^{\rm obs}$)\footnote{Average column  density, derived 
from the estimated mass and radius {\it before\/} deconvolution.} &
($\bar{N}_{H_2}^{\rm dec}$)\footnote{Average column  density, derived 
from the estimated mass and radius {\it after\/} deconvolution.} &
($n_{H_2}^{\rm pk}$)\footnote{Beam-averaged peak volume density at the 
resolution of the 500 $\mu$m data, derived from the peak column density 
on the basis of an assumed Gaussian source profile, using:
$ n_{{\rm H}_2}^{\rm pk}=\sqrt{4\ln2/\pi}\,\,N_{{\rm H}_2}^{\rm pk}\,\,/\,\,\bar{FWHM}_{500}$.} &
($\bar{n}_{H_2}^{\rm obs}$)\footnote{Average volume density, based on
the estimated radius {\it before\/} deconvolution,
using: $\bar{n}_{H_2}^{\rm obs}=M_{\rm c}/(\frac{4}{3}\pi R_{\rm c}^3 \mu m_H)$, 
where $\mu = 2.8$.} &

($\bar{n}_{H_2}^{\rm dec}$)\footnote{Average volume density, based on
the estimated radius {\it after\/} deconvolution, but otherwise using the same equation.} &

($\alpha_{\rm BE}$)\footnote{Bonnor-Ebert mass ratio, $\alpha_{\rm BE}=M_{\rm BE}/M_{\rm c}$ (see Section 4).} & Core & Comments \\

 & HGBS\_J$^*$ &  &  &  &  &  &  &  &  &  &  &  &  &  &  & & type\footnote{1=starless(unbound); 2=prestellar; 3=candidate prestellar; 4=dense core with embedded protostar.}  &  \\

 &  & [h m s] & [$^\circ\,'\,''$] & [pc] & [pc] & [$M_\odot$] & [$M_\odot$] & [K] & [K] &
[$10^{21}{\rm cm}\!^{-\!2}$] &
[$10^{21}{\rm cm}\!^{-\!2}$] &  
[$10^{21}{\rm cm}\!^{-\!2}$] &  
[$10^4{\rm cm}\!^{-\!3}$] &
[$10^4{\rm cm}\!^{-\!3}$] & 
[$10^4{\rm cm}\!^{-\!3}$] & & &  \\
 \hline
     1 & 040924.1+284723 & 04 09 24.10 &   +28 47 23 &   0.030 &  0.027 &   0.040 &  0.009 &    12.4 &   0.5 &    0.8 &   0.63 &  0.78 &    0.7 &   0.51 &   0.71 &    13.8 &       1 &  \\
    17 & 041042.8+281925 & 04 10 42.84 &   +28 19 25 &   0.017 &  0.015 & 0.061 & 0.027 &     8.9 &   0.7 &    2.2 &   3.13 &   3.73 &    2.2 &   4.55 &   5.93 &     3.7 &       3 &  \\
    76 & 041335.4+282117 & 04 13 35.45 &   +28 21 17 &   0.046 &  0.045 & 1.488 & 0.243 &     8.7 &   0.2 &   41.1 &   9.86 &  10.67 &   23.4 &   5.17 &   5.82 &     0.4 &       2 &  \\
   421 & 041840.7+281916 & 04 18 40.67 &   +28 19 16 &   0.013 &  0.010 &  0.026&  0.008 &    22.9 &   2.2 &    2.4 &   2.02 &   3.80 &    3.0 &   3.66 &   9.43 &    14.4 &       4 &  \\
\hline
\end{tabular}
\end{minipage}
\end{table}
\end{landscape}

\bsp

\label{lastpage}

\end{document}